\documentclass[10pt]{article}
\usepackage[utf8]{inputenc}
\usepackage{amssymb}
\usepackage{tabularx}
\usepackage{multirow}
\usepackage{float}
\usepackage{hyperref}
\usepackage{amsmath}
\usepackage{adjustbox}
\usepackage[table]{xcolor}
\usepackage{graphicx}
\usepackage{sectsty}
\usepackage{bbold}
\usepackage{wrapfig}
\usepackage{subfigure}
\usepackage{multirow}
\usepackage[labelfont=bf,textfont=md]{caption} 
\usepackage{amsthm}  % for mathematical definitions etc.
\usepackage[left=2cm,right=2cm,top=2.5cm,bottom=2.5cm]{geometry}
\usepackage{mathrsfs}
\usepackage[framemethod=tikz]{mdframed} % is used to build our "assumption" environment using \newmdtheoremenv
\usepackage{comment}
\usepackage{titling}
\usepackage{upgreek}
\usepackage{cite}

\usepackage{authblk} % for author and affiliation
\newcommand{\tens}[1]{%
  \mathbin{\mathop{\otimes}\limits_{#1}}%
}
\DeclareUnicodeCharacter{03C0}{\varnothing}
\DeclareUnicodeCharacter{0302}{\varnothing}

\begin{document}
\renewcommand{\thesection}{\Roman{section}} 
\renewcommand{\thesubsection}{\thesection.\Roman{subsection}}
\title{\textbf{Prediction of an $I(J^{P})=0(1^{-})$ $\bar{b}\bar{b}ud$ Tetraquark Resonance Close to the $B^\ast B^\ast$ Threshold Using Lattice QCD Potentials}}

\author[1]{Jakob Hoffmann \thanks{\href{mailto:jhoffmann@itp.uni-frankfurt.de}{jhoffmann@itp.uni-frankfurt.de}}}
\author[1,2]{Marc Wagner \thanks{\href{mailto:mwagner@itp.uni-frankfurt.de}{mwagner@itp.uni-frankfurt.de}}}
\affil[1]{\normalsize Institut f\"ur Theoretische Physik, Goethe Universit\"at Frankfurt am Main, Max-von-Laue-Stra{\ss}e 1, D-60438~Frankfurt~am~Main, Germany}
\affil[2]{\normalsize Helmholtz Research Academy Hesse for FAIR, Campus Riedberg, Max-von-Laue-Stra{\ss}e 12, D-60438~Frankfurt~am~Main, Germany }

\date{December 09, 2024}

% ------------------------------------------------------------

\maketitle

\begin{abstract}

We use antistatic-antistatic potentials computed with lattice QCD and a coupled-channel Born-Oppenheimer approach to explore the existence of a $\bar{b} \bar{b} u d$ tetraquark resonance with quantum numbers $I(J^P) = 0(1^-)$. A pole in the $\mbox{T}$ matrix signals a resonance with mass $m = 2 m_B + 94.0^{+1.3}_{-5.4} \, \text{MeV}$ and decay width $\Gamma = 140^{+86}_{-66} \, \text{MeV}$, i.e.\ very close to the $B^\ast B^\ast$ threshold. We also compute branching ratios, which clearly indicate that this resonance is mainly composed of a $B^\ast B^\ast$ meson pair with a significantly smaller $B B$ contribution. By varying the potential matrix responsible for the coupling of the $B B$ and the $B^\ast B^\ast$ channel as well as the $b$ quark mass, we provide additional insights and understanding concerning the formation and existence of the resonance.
We also comment on the importance of our findings and the main takeaways for a possible future full lattice QCD investigation of this $I(J^P) = 0(1^-)$ $\bar{b} \bar{b} u d$ tetraquark resonance.

\end{abstract}

%---------------------------------------------------------------------------------
%---------------------------------------------------------------------------------
%---------------------------------------------------------------------------------

\section{Introduction}\label{sec1}

There has been a long standing effort to experimentally find and theoretically predict exotic mesons, i.e.\ mesons with a more complicated structure than just a quark-antiquark pair. Tetraquarks, a specific class of exotic mesons, are color neutral states of two quarks and two antiquarks. They were first proposed almost 50 years ago in Ref.\ \cite{PhysRevD.15.267}.
A very clear experimental verification of the existence of tetraquarks was the detection of the electrically charged $Z_b$ and $Z_c$ states around 10 years ago (see Refs.\ \cite{Belle:2011aa,Xiao:2013iha,LHCb:2014zfx,BESIII:2013ouc,LHCb:2015sqg,BESIII:2015pqw}). Their masses and decay channels strongly suggest a heavy $\bar b b$ or $\bar c c$ pair, while their non-vanishing electrical charge implies the presence of an additional light quark-antiquark pair.
Recently, another very interesting and clear tetraquark system was found at LHCb, the $T_{cc}(3875)^+$ \cite{LHCb:2021vvq}. This state is composed of two heavy $c$ quarks and two light antiquarks and has quantum numbers $I(J^P) = 0(1^+)$. Its mass is located slightly below the lightest meson-meson threshold and it is, thus, almost QCD-stable.

From the theoretical perspective antiheavy-antiheavy-light-light ($\bar Q \bar Q q q$) systems (or equivalently heavy-heavy-antilight-antilight states) are particularly interesting, since they are somewhat simpler to study than some of the more common $\bar Q Q \bar q q$ systems, mainly because of two reasons: (1) quark-antiquark annihilation is not possible; (2) they can only decay into two heavy-light mesons, but not into quarkonium and a light meson. There are quite a number of papers studying such $\bar Q \bar Q q q$ tetraquarks based on a variety of phenomenological or QCD-based approaches including
quark models \cite{Carlson:1987hh,Manohar:1992nd,Chow:1994mu,Brink:1998as,Vijande:2003ki,Vijande:2006jf,Ebert:2007rn,Zhang:2007mu,Lipkin:2007cg,Godfrey:2008nc,Lee:2009rt,Karliner:2013dqa,Bicudo:2015bra,Karliner:2017qjm,Eichten:2017ffp,Park:2018wjk,Richard:2018yrm,Meng:2021yjr,Guo:2021yws,Richard:2022fdc,Ortega:2023pmr,Liu:2023vrk}
, sum rules \cite{Matheus:2007zz,Navarra:2007yw,Chen:2013aba,Wang:2017uld,Agaev:2016dsg,Tang:2019nwv,Agaev:2020dba,Agaev:2021vur}
, effective field theories \cite{Ohkoda:2012hv,Braaten:2014qka,Segovia:2016rwd,Soto:2020xpm,Braaten:2020nwp,Maiani:2022qze,Berwein:2024ztx,Mutuk:2023oyz}
and functional methods \cite{Heupel:2012ua,Santowsky:2021bhy,Hoffer:2024fgm}.

The first lattice QCD studies of $\bar Q \bar Q q q$ tetraquarks, mostly of the $\bar b \bar b u d$ tetraquark with quantum numbers $I(J^P) = 0(1^+)$, which is the bottom counterpart of the experimentally found $T_{cc}(3875)^+$, were based on antistatic-antistatic potentials \cite{PhysRevD.76.114503,Wagner:2010ad,Bali:2011gq,Brown:2012tm,Bicudo:2015kna}.
For heavy quark masses significantly larger than the typical QCD scale relevant for the dynamics of the light quarks, a $\bar Q \bar Q q q$ tetraquark system can be described by a non-relativistic Hamiltonian in the so-called Born-Oppenheimer approximation (see e.g.\ Ref.\ \cite{Braaten:2014qka}).
The first crude predictions of that type just used a single $\bar Q \bar Q$ potential from lattice QCD and a simple single-channel Schr\"odinger equation and, thus, did not distinguish between $B$ and $B^\ast$ mesons \cite{Bicudo:2012qt,Brown:2012tm,Bicudo:2015kna}. For example, Ref.\ \cite{Bicudo:2015kna} found a a binding energy of $90_{-43}^{+36} \, \text{MeV}$ for the $I(J^P) = 0(1^+)$ $\bar b \bar b u d$ tetraquark at physically light $u$ and $d$ quark masses. Later this approach was refined, by deriving a coupled-channel Schrödinger equation taking into account both an attractive and a repulsive $\bar Q \bar Q$ potential and the mass difference of $B$ and $B^\ast$. This inclusion of heavy spin effects resulted in a significantly reduced binding energy $59_{-38}^{+30} \, \text{MeV}$ \cite{Bicudo:2016ooe}, indicating the importance of proper multi-channel equations in the Born-Oppenheimer approach. For a recent and very comprehensive discussion and application of Born-Oppenheimer effective field theory to arbitrary XYZ exotic states we refer to Ref.\ \cite{Berwein:2024ztx}.

$\bar{b} \bar{b} u d$ tetraquarks with quantum numbers $I(J^P)=0(1^+)$, as well as related systems with flavor content $\bar{b} \bar{b} u s$, $\bar{b} \bar{c} u d$ and $\bar{c} \bar{c} u d$, have meanwhile also been investigated rigorously within full lattice QCD by several independent groups (see Refs.\
\cite{Francis:2016hui,Junnarkar:2018twb,Leskovec:2019ioa,Hudspith:2020tdf,Mohanta:2020eed,Padmanath:2022cvl,Meinel:2022lzo,Hudspith:2023loy,Aoki:2023nzp,Padmanath:2023rdu,Alexandrou:2023cqg,Alexandrou:2024iwi,Radhakrishnan:2024ihu,Collins:2024sfi,Whyte:2024ihh}). These computations point towards a binding energy of around $100 \, \text{MeV}$ for the $I(J^P)=0(1^+)$ $\bar{b} \bar{b} u d$ tetraquark, which is in fair agreement with the previously mentioned result $59_{-38}^{+30} \, \text{MeV}$ from the refined Born Oppenheimer approach \cite{Bicudo:2016ooe}.

Full lattice QCD computations with either $\bar{Q} \bar {Q} = \bar{b} \bar {b}$ or $\bar{Q} \bar {Q} = \bar{b} \bar {c}$ focus mainly on QCD-stable $\bar Q \bar Q q q$ tetraquarks \footnote{An exception is Ref.\ \cite{Alexandrou:2023cqg}, where indications for the existence of $\bar b \bar c u d$ tetraquark resonances were found.}. However, in addition to these stable states there might also exist short-lived tetraquark resonances.
For example in Ref.\ \cite{Bicudo:2017szl} a $\bar{b} \bar{b} u d$ tetraquark resonance with quantum numbers $I(J^P)=0(1^-)$ was predicted using lattice QCD potentials and the Born-Oppenheimer approach. There, a simple single-channel scattering problem was solved and a resonance was found $17^{+4}_{-4} \, \mathrm{MeV}$ above the lightest meson-meson threshold. Since in that work the $B$ and the $B^\ast$ mesons were treated as degenerate in mass, it could only be speculated, whether the predicted resonance should be expected near the $B B$ or the $B^\ast B^\ast$ threshold or somewhere in-between. In the present work we combine the refined coupled-channel Born-Oppenheimer approach from Ref.\ \cite{Bicudo:2016ooe} with scattering theory and thoroughly search for the $I(J^P)=0(1^-)$ $\bar{b} \bar{b} u d$ tetraquark resonance in the energy region between the $B B$ threshold and slightly above the $B^\ast B^\ast$ threshold.
Prelimary results obtained at an early stage of this project indicated that there is no resonance in the region close to the $BB$ threshold \cite{Hoffmann:2022jdx}. Motivated by the prediction in Ref. \cite{Bicudo:2017szl}, in that study we considered only energies up to $30\, \text{MeV}$ below the $B^{*}B^{*}$ threshold but not above the $B^{*}B^{*}$ threshold. In the following sections we give theoretical arguments, why the resonance crudely predicted in Ref.\ \cite{Bicudo:2017szl} should rather be expected near the $B^\ast B^\ast$ threshold, and present numerical results on the resonance mass confirming this expectation. Moreover, we compute branching ratios indicating that this resonance is mainly a $B^\ast B^\ast$ system with a significantly smaller $B B$ component.

This paper is structured as follows.
In Section~\ref{sec2} we review the computation of antistatic-antistatic potentials with lattice QCD. In Section~\ref{SEC003} we derive a $16 \times 16$ coupled-channel Schr\"odinger equation for these potentials. We also decompose this equation into smaller independent blocks corresponding to definite spin quantum numbers. In Section~\ref{SEC_scattering} we use standard techniques from non-relativistic scattering theory to relate the relevant coupled-channel Schr\"odinger equation to the elements of the $\mbox{T}$ matrix. In Section~\ref{SEC005} we solve the Schr\"odinger equation numerically and search for poles of the $\mbox{T}$ matrix in the complex energy plane. This provides the mass, the decay width and the $B B$ and $B^\ast B^\ast$ branching ratios of the $I(J^P)=0(1^-)$ $\bar{b} \bar{b} u d$ tetraquark resonance.
We also explore the dependence of our results on the potential matrix responsible for the coupling of the $B B$ and the $B^\ast B^\ast$ channel as well as on the heavy quark mass, to provide additional insights and understanding concerning the formation and existence of a resonance of that type.
We conclude in Section~\ref{SEC_conclusions}.

%---------------------------------------------------------------------------------
%---------------------------------------------------------------------------------
%---------------------------------------------------------------------------------

\section{$\bar{Q} \bar{Q} q q$ Static Potentials from Lattice QCD}\label{sec2}

The Born-Oppenheimer approach for antiheavy-antiheavy-light-light tetraquarks consists of two independent steps. In the first step one uses the static-quark approximation for the heavy antiquarks $\bar{Q} \bar{Q}$ and computes the corresponding antistatic-antistatic potentials in the presence of two light quarks $q q$ with lattice QCD. 
In this section we briefly review this first step to keep the paper self-contained (see Ref.\ \cite{Bicudo:2015kna} for more details) and introduce the relevant notation.
In the second step of the Born-Oppenheimer approach, one inserts the static potentials obtained in step~1 into coupled-channel Schr\"odinger equations to predict masses and properties of QCD-stable bound states or resonances. At this stage we also include the spin of the heavy $\bar{b}$ quarks. The main focus of this work is on step~2, for which we extend methods developed in Ref.\ \cite{Bicudo:2016ooe} for the QCD-stable $\bar b \bar b u d$ tetraquark with quantum numbers $I(J^P) = 0(1^+)$ to a $\bar b \bar b u d$ tetraquark resonance with quantum numbers $I(J^P) = 0(1^-)$.

In Refs.\ \cite{Wagner:2010ad,Bicudo:2015kna} antistatic-antistatic potentials were computed for isospin $I = 0, 1$. These potentials were extracted from the asymptotic $t$ behavior of Euclidean two-point correlation functions
\begin{equation}
	C(t) = \langle \Omega | \mathcal{O}^\dagger(\vec{r}_{1},\vec{r}_{2},t) \mathcal{O}(\vec{r}_{1},\vec{r}_{2},0) | \Omega \rangle .
\end{equation}
In this work we are exclusively interested in $I = 0$, for which the relevant operators are
\begin{equation}
	\mathcal{O}(\vec{r}_{1},\vec{r}_{2}) = (\mathbf{C}\mathbb{L})_{AB} (\mathbf{C}\mathbb{S})_{CD} \Big(\bar{Q}^{a}_C(\vec{r}_{1}) u^{a}_A(\vec{r}_{1})\Big)\Big(\bar{Q}^{b}_D(\vec{r}_{2})d^{b}_B(\vec{r}_{2})\Big) - (u \leftrightarrow d) ,
	\label{eq:0002}
\end{equation}
where $\vec{r}_{1}$ and $\vec{r}_{2}$ denote the positions of the static antiquarks $\bar{Q}$, $\mathbf{C}=\gamma_{0} \gamma_{2}$ is the charge conjugation matrix, $a,b = 1,2,3$ are color indices, $A,B,C,D = 1,2,3,4$ are spin indices, $\mathbb{L}$ is the light quark spin matrix and $\mathbb{S}$ is the heavy quark spin matrix.

There are just 4 linearly independent possibilities for the heavy quark spin matrix,
\begin{equation}
	\mathbb{S} \in \{(\mathbb{1} + \gamma_{0}) \gamma_{5} , (\mathbb{1} + \gamma_{0}) \gamma_{j} \}
	\label{EQN_S}
\end{equation}
with $j = 1,2,3$, because static quark spinors have only two components. Since static potentials are degenerate with respect to the heavy quark spins, the concrete choice of $\mathbb{S}$ is irrelevant in this section. It will, however, become important in Section~\ref{SEC003}, when we set up coupled-channel Schr\"odinger equations and distinguish $B$ and $B^\ast$ mesons.

For the light quark spin matrix there are 16 linearly independent possibilities, $\mathbb{L} \in \{
(\mathbb{1} \pm \gamma_{0}) \gamma_{5} , 
(\mathbb{1} \pm \gamma_{0}) \gamma_{j} ,
(\mathbb{1} \pm \gamma_{0}) \mathbb{1} ,
(\mathbb{1} \pm \gamma_{0}) \gamma_{5} \gamma_{j}
\}$,
chosen such that the corresponding trial states $\mathcal{O}^{\dagger}(\vec{r}_{1},\vec{r}_{2}) |\Omega \rangle$ have definite quantum numbers $|j_z|$, $\mathcal{P}$ and $\mathcal{P}_x$. $|j_z|$ is the total angular momentum of the light quarks and gluons in the direction of the $\bar{Q} \bar{Q}$ separation axis (for simplicity one can choose the $z$ axis), $\mathcal{P}$ denotes parity and $\mathcal{P}_x$ the behavior under reflection along an axis perpendicular to the separation axis (e.g.\ the $x$ axis).
At large $\bar Q \bar Q$ separations the $\bar Q \bar Q q q$ system will consist of two heavy light mesons, either $B$, $B^\ast$, $B_0^\ast$ and/or $B_1^\ast$. In this work we are only interested in the negative parity mesons $B$ and $B^\ast$, but not their positive parity partners $B_0^\ast$ and/or $B_1^\ast$, which are around $500 \, \text{MeV}$ heavier. Thus, we just need to consider potentials with asymptotic values of two times the $B$ and/or $B^\ast$ meson mass ($m_B = m_{B^\ast}$ for static heavy quarks), but not with higher asymtotic values corresponding to a negative and a positive parity meson or two positive parity mesons. One can show that the relevant choices for the light quark spin matrix are then limited to
\begin{equation}
	\mathbb{L} \in \{ (\mathbb{1} + \gamma_{0}) \gamma_{5} , (\mathbb{1} + \gamma_{0}) \gamma_{j} \}
	\label{eq: 0006}
\end{equation}
(see Ref.\ \cite{Bicudo:2015kna} for details), which lead to two different potentials. $\mathbb{L} = (\mathbb{1} + \gamma_{0}) \gamma_{5}$ corresponds to a strongly attractive potential, denoted as $V_5(r)$, whereas $\mathbb{L} = (\mathbb{1} + \gamma_{0}) \gamma_{j}$ corresponds to a significantly weaker repulsive potential, denoted as $V_j(r)$ ($r = |\vec{r}_{2}-\vec{r}_{1}|$ is the $\bar Q \bar Q$ separation).

\begin{figure}[H]
	\centering
	\includegraphics[scale=0.53]{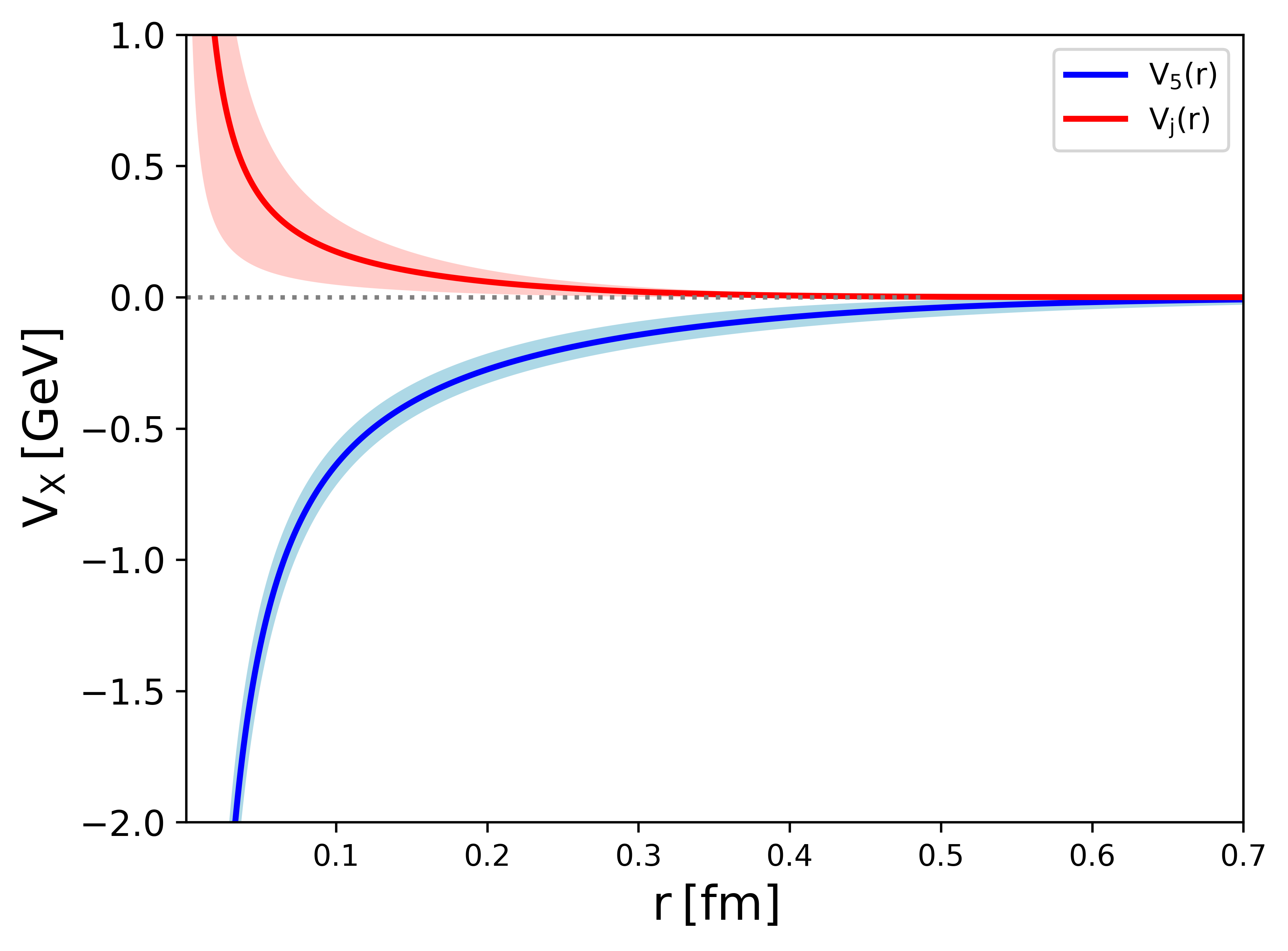}
	\caption{Potential parameterizations $V_5$ and $V_j$ of lattice QCD results from Ref.\ \cite{Bicudo:2015kna} as functions of the $\bar Q \bar Q$ separation $r$.}
	\label{fig:static-potentials}
\end{figure}

The lattice QCD results for $V_{5}(r)$ and $V_{j}(r)$ from Ref.\ \cite{Bicudo:2015kna} (with two times the lightest static-light meson mass subtracted), extrapolated to the physical pion mass $m_{\pi} \approx 140 \, \text{MeV}$, can be consistently parameterized by
\begin{equation}
	V_{X}(r) = -\frac{\alpha_{X}}{r} e^{-(r/d_X)^2} \quad , \quad X = 5,j
	\label{eq:0007}
\end{equation}
with parameters $\alpha_{5} = 0.34 \pm 0.03$, $d_{5}=0.45^{+0.12}_{-0.10} \, \text{fm}$, $\alpha_{j} = -0.10 \pm 0.07$ and $d_{j}=(0.28 \pm 0.017) \text{fm}$ determined by $\chi^2$-minimizing fits in Refs.\ \cite{Bicudo:2015kna,Bicudo:2016ooe}. These potential parameterizations are shown in Figure~\ref{fig:static-potentials}. For small $\bar Q \bar Q$ separations the potentials are dominated by 1-gluon exchange between the static quarks and, thus, are proportional to $1/r$. For large separations, the potentials are exponentially screened and represent rather weak residual meson-meson interactions.

%---------------------------------------------------------------------------------
%---------------------------------------------------------------------------------
%---------------------------------------------------------------------------------

\section{\label{SEC003}Coupled-Channel Schr\"odinger Equation for $I(J^P) = 0(1^-)$}

%---------------------------------------------------------------------------------

\subsection{$16\times 16$ Coupled-Channel Schr\"odinger Equation}

The potentials $V_5(r)$ and $V_j(r)$ discussed in Section~\ref{sec2} are independent of the static quark spins. Now we include heavy quark spin effects in our approach by setting the asymptotic potential values to the corresponding two meson thresholds, $2 m_B$, $m_B + m_{B^\ast}$ and $2 m_{B^\ast}$, respectively. We take the meson masses $m_B$ and $m_{B^\ast}$ from experiments \cite{ParticleDataGroup:2024cfk}, most importantly the mass difference $m_{B^\ast} - m_B = 45 \, \text{MeV}$.

Following Ref.\ \cite{Bicudo:2016ooe} a $16 \times 16$ Hamiltonian can be defined as
\begin{equation}
	H = H_{0} + H_{\text{int}}
    \label{eq:TotalHamiltonian}
\end{equation}
with a free part
\begin{equation}
	H_{0} = M \tens{} \mathbb{1}_{4\times 4} + \mathbb{1}_{4\times 4} \tens{} M + \frac{\vec{p}_{1}^{\,2} + \vec{p}_{2}^{\,2}}{2m_{b}}
\end{equation}
essentially describing non-interacting $B$ and $B^\ast$ mesons. $\vec{p}_{1}$ and $\vec{p}_{2}$ are the momenta of the heavy $\bar{b}$ quarks, $m_{b}$ denotes the $b$ quark mass and $M = \text{diag}(m_{B},m_{B^{*}},m_{B^{*}},m_{B^{*}})$ is a $4 \times 4$ diagonal matrix containing the masses of $B$ and $B^{*}$ mesons. The corresponding free Schr\"odinger equation in the center of mass frame is a partial differential equation for the relative coordinate of the $\bar{b}$ quarks $\vec{r} = \vec{r}_2 - \vec{r}_1$,
\begin{equation}
	H_0 \vec{\Psi}(\vec{r}) = E \vec{\Psi}(\vec{r}) \quad , \quad H_{0} = M \tens{} \mathbb{1}_{4\times 4} + \mathbb{1}_{4\times 4} \tens{} M + \frac{\vec{p}^{\,2}}{2 \mu}
	\label{eq:0008} 
\end{equation}
with the reduced mass $\mu = m_b / 2$. The wave function $\vec{\Psi}(\vec{r})$ has 16 components, where each component is associated with a specific meson-meson pair,
\begin{eqnarray}
	\nonumber & & \hspace{-0.7cm} \vec{\Psi} \equiv (
	B B     , B B_x^*     , B B_y^*     , B B_z^*     \ \ , \ \
	B_x^* B , B_x^* B_x^* , B_x^* B_y^* , B_x^* B_z^* \ \ , \ \
	B_y^* B , B_y^* B_x^* , B_y^* B_y^* , B_y^* B_z^* \ \ , \ \ \\
	\label{EQN003} & & \hspace{0.675cm} B_z^* B , B_z^* B_x^* , B_z^* B_y^* , B_z^* B_z^*)^T
\end{eqnarray}
with the indices $x,y,z$ denoting the three possible spin orientations of a $B^{*}$ meson.

The interacting part $H_\text{int}$ contains the potentials $V_5(r)$ and $V_j(r)$. It is a $16\times 16$ non-diagonal matrix, which couples the 16 independent partial differential equations from Eq.\ (\ref{eq:0008}). This is so, because the potentials $V_5(r)$ and $V_j(r)$ do not correspond to simple combinations of two specific mesons, but rather to linear combinations of several $B$ and/or $B^\ast$ meson pairs as we work out in detail in the following. To derive the mathematical structure of $H_\text{int}$, we use
\begin{equation}
	H_\text{int} = T^{-1} V_\text{diag}(r) T \quad , \quad V_\text{diag} = \text{diag}(\underbrace{V_{5}(r), \dots ,V_{5}(r)}_{4\times}, \underbrace{V_{j}(r), \ldots ,V_{j}(r)}_{12\times})
	\label{eq: Interaction-Hamiltonian} 
\end{equation}
as a starting point. $T$ is a $16\times 16$ transformation matrix, whose entries are determined by rewriting the interpolating operators defined in equation \eqref{eq:0002} in terms of combinations of heavy-light mesons. In other words, $T$ transforms the 16 operators (\ref{eq:0002}) with $\mathbb{S}$ and $\mathbb{L}$ chosen according to Eqs.\ (\ref{EQN_S}) and (\ref{eq: 0006}) to the 16 meson-meson combinations defined in Eqs.\ (\ref{eq:0008}) and (\ref{EQN003}) via the Schr\"odinger equation and the corresponding 16-component wave function. The entries of $T$ can be obtained by expressing the operators (\ref{eq:0002}) in terms of quark-antiquark combinations, not only in color space, but also in spin space, using Fierz identities,
\begin{equation}
	\mathcal{O}(\vec{r}_{1},\vec{r}_{2}) = \mathbb{G}(\mathbb{S},\mathbb{L})_{AB} \Big(\bar{Q}(\vec{r}_{1}) \Gamma_A u(\vec{r}_{1})\Big) \Big(\bar{Q}(\vec{r}_{2}) \Gamma_B d(\vec{r}_{2})\Big) - (u \leftrightarrow d)
	\label{EQN011} 
\end{equation}
with coefficients 
\begin{equation}
	\mathbb{G}(\mathbb{S},\mathbb{L})_{AB} = \frac{1}{16} \text{Tr} \Big((\mathbf{C}\mathbb{S})^{T} \Gamma_{A}^T (\mathbf{C}\mathbb{L}) \Gamma_{B}\Big) .
	\label{eq:Fierz-coefficients}
\end{equation}
The implicit sums over $A$ and $B$ in Eq.\ (\ref{EQN011}) are over four matrices each, i.e.\ $\Gamma_A , \Gamma_B \in \{
(\mathbb{1} + \gamma_{0}) \gamma_{5} ,
(\mathbb{1} + \gamma_{0}) \gamma_{j} \}$, where $\bar{Q} \Gamma_A q$ with $\Gamma_A = (\mathbb{1} + \gamma_{0}) \gamma_{5}$ represents a $B$ meson (quantum numbers $J^P = 0^-$) and $\bar{Q} \Gamma_A q$ with $\Gamma_A = (\mathbb{1} + \gamma_{0}) \gamma_{j}$ represents one of the three possible spin orientations of a $B^\ast$ meson (quantum numbers $J^P = 1^-$). The entries of the matrix $T$ are given by the coefficients $\mathbb{G}(\mathbb{S},\mathbb{L})_{AB}$:
\begin{itemize}
\item The row of $T$ is given by the choice of $\mathbb{S}$ and $\mathbb{L}$ and a mapping consistent with the right equation in (\ref{eq: Interaction-Hamiltonian}). This requires that rows 1 to 4 correspond to $\mathbb{L} = (\mathbb{1} + \gamma_{0}) \gamma_{5}$ and rows 5 to 16 to $\mathbb{L} = (\mathbb{1} + \gamma_{0}) \gamma_{j}$. The mapping with respect to $\mathbb{S}$ is not unique. For example, one can define that rows 1, 5, 9, 13 correspond to $\mathbb{S} = (\mathbb{1} + \gamma_{0}) \gamma_{5}$, rows 2, 6, 10, 14 correspond to $\mathbb{S} = (\mathbb{1} + \gamma_{0}) \gamma_1$, etc.

\item The column of $T$ is given by the indices $A$ and $B$ and the mapping (\ref{EQN003}).
\end{itemize}

In total, we now have the Schr\"odinger equation
\begin{equation}
	H \vec{\Psi}(\vec{r}) = \Big(H_0 + H_\text{int}\Big) \vec{\Psi}(\vec{r}) = E \vec{\Psi}(\vec{r})
	\label{EQN_SE}
\end{equation}
with $H_0$ from Eq.\ (\ref{eq:0008}) and $H_\text{int}$ from Eq.\ (\ref{eq: Interaction-Hamiltonian}). This is a coupled-channel partial differential equation in the relative coordinate $\vec{r}$. The 16 equations are coupled by the non-vanishing off-diagonal elements of $H_\text{int}$.

%---------------------------------------------------------------------------------

\subsection{Radial $1 \times 1$ and $2 \times 2$ Coupled-Channel Schr\"odinger Equations}\label{sec:symmetries}

Since there is no spin-orbit coupling in our Hamiltonian, the orbital angular momentum operators $\vec{L}^2$ and $L_z$ commute with $H$. Thus, $L$ and $L_z$, representing the relative orbital angular momentum of the two $\bar b$ quarks, can be used as quantum numbers. This allows to write the wave function as a product of a radial wave function and a spherical harmonic function, $\vec{\Psi}(\vec{r}) = \vec{\varphi}_L(r) Y_{L,L_z}(\theta,\phi)$, and to simplify the Schr\"odinger Equation (\ref{EQN_SE}) to a radial Schr\"odinger Equation, i.e.\ an ordinary differential equation in the $\bar b \bar b$ separation $r = |\vec{r}|$.

We denote the spin of both, the light and the heavy quarks, represented by the 16 components of the wave function as specified in Eq.\ (\ref{EQN003}), by $\vec{S}$. One can show that $\vec{S}^2$ and $S_z$ commute with $H$, $\vec{L}^2$ and $L_z$. Thus, in addition to $L$ and $L_z$ one can also use $S$ and $S_z$ as quantum numbers. As a consequence, one can decouple the $16 \times 16$ radial Schr\"odinger equation into smaller blocks labeled by $S$ and $S_z$. Standard spin coupling of four spin $1/2$ particles is given by $\mathbf{2} \tens{} \mathbf{2} \tens{} \mathbf{2} \tens{} \mathbf{2} = \mathbf{2} \oplus \mathbf{9} \oplus \mathbf{5}$, where $\mathbf{2}$ on the right hand side denotes a $2 \times 2$ block with $S = 0$, $\mathbf{9}$ denotes 3 degenerate $3 \times 3$ blocks with $S = 1$ and $\mathbf{5}$ denotes 5 degenerate $1 \times 1$ blocks with $S = 2$. The decoupling was already worked out in Ref.\ \cite{Bicudo:2016ooe}. In the following we just summarize the resulting equations.

% *****

\subsubsection*{\label{SEC_radialSE}Radial Schr\"odinger Equation for $S = 0$}

For $S=0$ there is a single $2\times 2$ equation,
\begin{equation}
	\bigg(\begin{pmatrix} 2 m_B & 0 \\ 0 & 2 m_{B^\ast} \end{pmatrix} - \frac{\nabla^2}{2 \mu} + H_{\text{int}, S=0}\bigg) \vec{\varphi}_{L,S=0}(r) = E \vec{\varphi}_{L,S=0}(r)
	\label{eq:spin0-equation}
\end{equation}
with
\begin{equation}
	\nabla^2 = \frac{d^2}{d r^2} + \frac{2}{r} \frac{d}{d r} - \frac{L (L+1)}{r^2}
	\label{EQN_nabla_sq}
\end{equation}
and
\begin{equation}
	H_{\text{int}, S=0} = \frac{1}{4} \begin{pmatrix} V_{5}(r) + 3 V_{j}(r) & \sqrt{3} (V_{5}(r) - V_{j}(r)) \\ \sqrt{3} (V_{5}(r) - V_{j}(r)) & 3 V_{5}(r) + V_{j}(r) \end{pmatrix} .
	\label{eq:potential-matrix-2x2}
\end{equation}
The 2 components of the wave function represent the following meson-meson combinations:
\begin{equation}
	\vec{\varphi}_{L,S=0} \equiv \bigg( BB \ \ , \ \ \frac{1}{\sqrt{3}} \vec{B}^\ast \vec{B}^\ast \bigg)^T = \bigg( BB \ \ , \ \ \frac{1}{\sqrt{3}} \Big(B_x^\ast B_x^\ast + B_y^\ast B_y^\ast + B_z^\ast B_z^\ast\Big)\bigg)^T .
	\label{EQN_phi}
\end{equation}

% *****

\subsubsection*{Radial Schr\"odinger Equations for $S = 1$}

There are 3 identical $3 \times 3$ equations for $S = 1$ corresponding to $S_z = -1, 0, +1$. Each of these $3 \times 3$ equations can be decoupled further, into a $1 \times 1$ equation (i.e.\ a single equation), which is symmetric under meson exchange, and a $2 \times 2$ equation, which is antisymmetric under $B/B^\ast$ meson exchange.

The $1 \times 1$ equation is
\begin{equation}
	\bigg(m_B + m_B^\ast - \frac{\nabla^2}{2 \mu} + H_{\text{int}, S=1}^{(1 \times 1)}\bigg) \varphi_{L,S=1,S_z}^{(1 \times 1)}(r) = E \varphi_{L,S=1,S_z}^{(1 \times 1)}(r)
	\label{eq:spin1-equations-symmetric}
\end{equation}
with
\begin{equation}
	H_{\text{int}, S=1}^{(1 \times 1)} = V_j(r) \quad , \quad \varphi_{L,S=1,S_z}^{(1 \times 1)}(r) \equiv \frac{1}{\sqrt{2}} \Big(B_{S_z}^\ast B + B B_{S_z}^\ast\Big) .
\end{equation}

The $2 \times 2$ equation is
\begin{equation}
	\bigg(\begin{pmatrix} m_B + m_{B^\ast} & 0 \\ 0 & 2 m_{B^\ast} \end{pmatrix} - \frac{\nabla^2}{2 \mu} + H_{\text{int}, S=1}^{(2 \times 2)}\bigg) \vec{\varphi}_{L,S=1,S_z}^{(2 \times 2)}(r) = E \vec{\varphi}_{L,S=1,S_z}^{(2 \times 2)}(r)
	\label{eq:spin1-equation-antisymmetric}
\end{equation}
with
\begin{equation}
	H_{\text{int}, S=1}^{(2 \times 2)} = \frac{1}{2} \begin{pmatrix} V_5(r) + V_j(r) & V_5(r)-V_j(r) \\ V_5(r) - V_j(r) & V_5(r) + V_j(r) \end{pmatrix} \quad , \quad \vec{\varphi}_{L,S=1,S_z}^{(2 \times 2)}(r) \equiv \bigg( \frac{1}{\sqrt{2}} \Big(B_{S_z}^\ast B -B B_{S_z}^\ast\Big) \ \ , \ \ T_{1,S_z}(\vec{B}^\ast , \vec{B}^\ast)\bigg)^T ,
\end{equation}
where $T_{1,S_z}$ is a spherical tensor representing the coupling of two $B^\ast$ mesons to $S = 1$ with $z$ component $S_z$.

% *****

\subsubsection*{Radial Schr\"odinger Equations for $S = 2$}

There are 5 identical $1 \times 1$ equations for $S = 2$ corresponding to $S_z = -2, -1, 0, +1, +2$,
\begin{equation}
	\bigg(2 m_{B^\ast} - \frac{\nabla^2}{2 \mu} + H_{\text{int}, S=2}\bigg) \varphi_{L,S=2,S_z}(r) = E \varphi_{L,S=2,S_z}(r)
	\label{eq:spin2-equation}
\end{equation}
with
\begin{equation}
	H_{\text{int}, S=2} = V_j(r) \quad , \quad \varphi_{L,S=2,S_z}(r) \equiv T_{2,S_z}(\vec{B}^\ast , \vec{B}^\ast) ,
\end{equation}
where $T_{2,S_z}$ is a spherical tensor representing the coupling of two $B^\ast$ mesons to $S = 2$ with $z$ component $S_z$.

%---------------------------------------------------------------------------------

\subsection{Selecting the Coupled-Channel Schr\"odinger Equation for $I(J^P) = 0(1^-)$} 

The formalism we developed so far in this section and the resulting coupled-channel Schr\"odinger equations (\ref{eq:spin0-equation}), (\ref{eq:spin1-equations-symmetric}), (\ref{eq:spin1-equation-antisymmetric}) and (\ref{eq:spin2-equation}) are applicable to $\bar b \bar b u d$ systems with arbitrary quantum numbers $I$ and $L$. Since our goal in this work is a refined study of the $I(J^P) = 0(1^-)$ tetraquark resonance predicted in Ref.\ \cite{Bicudo:2017szl}, which has orbital angular momentum $1$, we fix $L = 1$, in particular in Eq.\ (\ref{EQN_nabla_sq}), for the remainder of this work.

In the following we discuss, which of the four equations (\ref{eq:spin0-equation}), (\ref{eq:spin1-equations-symmetric}), (\ref{eq:spin1-equation-antisymmetric}) and (\ref{eq:spin2-equation}) has to be selected to study the $I(J^P) = 0(1^-)$ tetraquark resonance from Ref.\ \cite{Bicudo:2017szl} in a refined way with effects due to the heavy quark spins and the corresponding $B$ and $B^\ast$ mass splitting included. We start by collecting all combinations of isospin $I$, light quark spin $S_q$, light color, heavy color, orbital angular momentum of the heavy quarks $L = 1$, heavy quark spin $S_Q$, total spin $S$ (with $\vec{S} = \vec{S}_q + \vec{S}_Q$) and total angular momentum $J$ (with $\vec{J} = \vec{L} + \vec{S}$) allowed by the Pauli principle and by color confinement of QCD (see Table~\ref{tab:summary-qn-odd-l}). There are two possibilities for isospin, $I = 0, 1$, as well as for light quark spin, $S_q = 0, 1$ (column 1 and 2). Moreover, we fix $L = 1$ (column 5), as already stated. The remaining entries in that table are consequences of the Pauli principle and of color confinement as discussed in the following items.
\begin{itemize}
\item color ($q q$): \\
Since the light quarks are fermions, they must form an antisymmetric combination. Thus, symmetry or antisymmetry with respect to $I$ and to $S_q$ fix the light color to either an antisymmetric $\bar 3$ or a symmetric $6$. For example, an antisymmetric $I = 0$ and an antisymmetric $S_q = 0$ imply an antisymmetric light color $\bar{3}$ (first line in Table~\ref{tab:summary-qn-odd-l}).

\item color ($\bar b \bar b$): \\
Color confinement requires a color singlet for the four quarks. Thus, if color ($q q$) is $\bar 3$, color ($\bar b \bar b$) must be $3$. If color ($q q$) is $6$, color ($\bar b \bar b$) must be $\bar 6$.

\item $S_Q$: \\
Since the $\bar b$ quarks are fermions, they must form an antisymmetric combination. Thus, symmetry or antisymmetry with respect to color ($\bar b \bar b$) and to $L = 1$ fix $S_Q$. For example, an antisymmetric $\text{color (}\bar b \bar b\text{)} = 3$ and an antisymmetric $L = 1$ imply an antisymmetric heavy quark spin $S_Q = 0$ (first line in Table~\ref{tab:summary-qn-odd-l}).

\item $S$: \\
$S$ is the usual coupling of $S_q$ and $S_Q$ with no further restrictions. The values of $S$ allow to assign the radial Schr\"odinger equations (\ref{eq:spin0-equation}), (\ref{eq:spin1-equations-symmetric}), (\ref{eq:spin1-equation-antisymmetric}) and (\ref{eq:spin2-equation}) to the four lines of Table~\ref{tab:summary-qn-odd-l} (last column ``equations''). Since $S = 1$ appears only once for $I = 0$, but twice for $I = 1$, it is clear that the $1 \times 1$ equation (\ref{eq:spin1-equations-symmetric}) corresponds to $I = 0$, $S = 1$ and the  $2 \times 2$ equation (\ref{eq:spin1-equation-antisymmetric}) to $I = 1$, $S = 1$. For the $2 \times 2$ equations we also note, whether a line in Table~\ref{tab:summary-qn-odd-l} corresponds to the attractive potential $V_5(r)$ or the repulsive potential $V_j(r)$. This is directly related to color ($\bar b \bar b$) and 1-gluon exchange dominating at small separations, where $\bar 3$ corresponds to an attractive potential and $6$ to a repulsive potential (see e.g.\ Ref.\ \cite{Bicudo:2015kna} for a detailed discussion).

\item $J^P$: \\
$J$ is the usual coupling of $L$ and $S$ with no further restrictions. Parity $P = -$ in all cases, because both the heavy quarks and the light quarks are in negative parity combinations (see Eqs.\ (\ref{EQN_S}) and (\ref{eq: 0006})) and orbital angular momentum $L = 1$ also contributes a factor of $-1$ to $P$.
\end{itemize}

\begin{table}[htb]
\begin{center}
\renewcommand{\arraystretch}{1.5}
\begin{tabular}{|c|c|c||c|c|c||c|c||c|}
    \hline
         \multicolumn{3}{|c||}{light quarks $qq$ } & \multicolumn{3}{c||}{heavy quarks $\bar{b}\bar{b}$} & \multicolumn{2}{c||}{tetraquark $\bar{b}\bar{b}qq$} & \multirow{2}{*}{equations} \\
    % \hline
    \cline{1-8}
         $I$ & $S_q$ & color ($q q$) & color ($\bar b \bar b$) & $L$ & $S_Q$ & $S$ & $J^{P}$ & \\
    \hline
         \multirow{2}{*}{ $0(A)$ } & $0(A)$ & $\bar{3}(A)$ & $3(A)$ & \multirow{6}{*}{ $1(A)$ } & $0(A)$ & $0$ & $1^{-}$ & $V_5(r)$ in Eq.\ (\ref{eq:spin0-equation}) \\
    \cline{2-4}\cline{6-9}
           & \multirow{3}{*}{ $1(S)$ } & \multirow{3}{*}{ $6(S)$ } & \multirow{3}{*}{ $\bar{6}(S)$ } & & \multirow{3}{*}{ $1(S)$ } & \multirow{3}{*}{ $0,1,2$ } & \multirow{3}{*}{ $0^{-},1^{-},2^{-},3^{-}$ } & $V_j(r)$ in Eq.\ (\ref{eq:spin0-equation}) \\
          & & & & & & & & Eq.\ (\ref{eq:spin1-equations-symmetric}) \\
          & & & & & & & & Eq.\ (\ref{eq:spin2-equation}) \\
    \cline{1-4}\cline{6-9}
         \multirow{2}{*}{ $1(S)$ } & $0(A)$ & $6(S)$ & $\bar{6}(S)$ & & $1(S)$ & $1$ & $0^{-},1^{-},2^{-}$ & $V_j(r)$ in Eq.\ (\ref{eq:spin1-equation-antisymmetric}) \\
    \cline{2-4}\cline{6-9}
          & $1(S)$ & $\bar{3} (A)$ & $3(A)$ &  & $0(A)$ & $1$ & $0^{-},1^{-},2^{-}$ & $V_5(r)$ in Eq.\ (\ref{eq:spin1-equation-antisymmetric}) \\
    \hline
\end{tabular}
\end{center}
\caption{Combinations of quantum numbers for $\bar{b} \bar{b} u d$ tetraquark systems with orbital angular momentum $L = 1$ of the two $\bar b$ quarks and the corresponding radial Schr\"odinger equations derived in Section~\ref{SEC_radialSE}. ``(S)'' and ``(A)'' denote symmetric and antisymmetric combinations, respectively.}
\label{tab:summary-qn-odd-l}
\end{table}

Now it is almost obvious, which of the radial coupled-channel equations (\ref{eq:spin0-equation}), (\ref{eq:spin1-equations-symmetric}), (\ref{eq:spin1-equation-antisymmetric}) and (\ref{eq:spin2-equation}) is appropriate to explore the $I(J^P) = 0(1^-)$ tetraquark resonance predicted in Ref.\ \cite{Bicudo:2017szl} in a refined way. Since it has $I = 0$, the last two lines of Table~\ref{tab:summary-qn-odd-l} can be discarded. Moreover, the equation must contain the attractive potential $V_5(r)$. The only candidate left is the $2 \times 2$ equation (\ref{eq:spin0-equation}) for $S = 0$. This equation, containing both the attractive $V_5(r)$ and the repulsive $V_j(r)$, is the generalization of the simple $1 \times 1$ equation used in Ref.\ \cite{Bicudo:2017szl}, which solely containes $V_5(r)$, and where heavy spin effects were ignored. From now on, we will exclusively focus on Eq.\ (\ref{eq:spin0-equation}).

We close this section by noting that Table~\ref{tab:summary-qn-odd-l} is essentially valid for arbitrary odd $L$. One just has to add $L-1$ to the entries in column ``$L$'' and column ``$J^P$''. For even $L$ one can follow the same logical steps, which will lead to a somewhat different table already presented and discussed in Ref.\ \cite{Bicudo:2016ooe} (Table~I in that reference). Moreover, notice that for $I = 1$ the potentials are different compared to the $I = 0$ case we are focusing on: $V_5(r)$ is repulsive and $V_j(r)$ is attractive, however, significantly less attractive than $V_5(r)$ for $I = 0$, because of the interaction of the two light spins (see Ref.\ \cite{Bicudo:2024vxq}).

In principle, one could study the existence of $\bar b \bar b u d$ bound states and resonances with arbitrary $I(J^P)$ quantum numbers within our approach using Table~I from Ref.\ \cite{Bicudo:2016ooe} and Table~\ref{tab:summary-qn-odd-l} from this work. However, states with quantum numbers different from $I(J^P) = 0(1^+)$ (and $L = 0$) and $I(J^P)=0(1^-)$ (and $L = 1$) are not expected to host resonances or bound states as explained in the following. As already noted, for Isospin $I=0$ only Eq. \eqref{eq:spin0-equation} with spin $S=0$ contains an attractive potential, whereas equations \eqref{eq:spin1-equations-symmetric} and \eqref{eq:spin2-equation} contain exclusively a repulsive potential. For $L=1$ this leaves the state with quantum numbers $I(J^P)=0(1^-)$ as the only candidate for a resonance with negative parity. The next candidate with negative parity would have orbital angular momentum $L=3$. Similarly, the next candidate for a resonance with positive parity would need to have at least $L=2$. Since cases with $L \geq 2$ did not show a clear signal for a resonance even in the single-channel approximation used in Ref.\ \cite{Bicudo:2017szl} (where repulsive interactions are neglected), we do not expect resonances in our refined coupled-channel approach. States with isospin $I=1$ are also not expected to host bound states or resonances due to the significantly less attractive potential \cite{Bicudo:2015kna}.

%---------------------------------------------------------------------------------
%---------------------------------------------------------------------------------
%---------------------------------------------------------------------------------

\section{\label{SEC_scattering}Scattering Formalism}

Now we prepare the coupled-channel Schr\"odinger equation (\ref{eq:spin0-equation}), describing possibly existing $\bar b \bar b u d$ tetraquark states with $I(J^P) = 0(1^-)$, for numerical scattering analyses, by applying basic techniques from scattering theory. For single-channel equations these techniques are extensively discussed in standard textbooks on quantum mechanics. For a generalization and discussion in the context of coupled-channel equations we refer e.g.\ to Refs.\ \cite{Bicudo:2019ymo,Bicudo:2020qhp,Bicudo:2022ihz}.

%---------------------------------------------------------------------------------

\subsection{Partial Wave Expansion}

The radial Schr\"odinger equation (\ref{eq:spin0-equation}) has two coupled-channels, a $B B$ channel and a $B^\ast B^\ast$ channel (see Eq.\ (\ref{EQN_phi})). The corresponding scattering momenta are
\begin{equation}
	k_{B B} = \sqrt{2 \mu (E - 2m_B)} \quad , \quad k_{B^\ast B^\ast} = \sqrt{2 \mu (E - 2m_{B^\ast})} .
	\label{eq:coupled-channel-momenta}
\end{equation}
Both momenta are real for energies $E > 2 m_{B^\ast}$, i.e.\ above the $B^\ast B^\ast$ threshold. For $2 m_B < E < 2 m_{B^\ast}$ only the $B B$ channel is open and $k_{B^\ast B^\ast}$ is purely imaginary. For $E < 2 m_B$, both channels are closed, scattering cannot take place and Eq.\ (\ref{eq:spin0-equation}) describes possibly existing bound states.

These three cases can be treated in the same way, by decomposing the wave function $\vec{\Psi}(\vec{r})$ into incident plane waves and emergent spherical waves and by carrying out a partial wave expansion. For $L = 1$ the radial wave function $\vec{\varphi}_{1,S=0}(r)$ appearing in Eqs.\ (\ref{eq:spin0-equation}) and (\ref{EQN_phi}) is then given by
\begin{equation}
	\vec{\varphi}_{1,S=0}(r) = \begin{pmatrix}
	A_{B B} j_1(k_{B B} r) + \chi_{B B}(r) / r \\
	A_{B^\ast B^\ast} j_1(k_{B^\ast B^\ast} r) + \chi_{B^\ast B^\ast}(r) / r
	\end{pmatrix} .
	\label{EQN_phi_decomposition}
\end{equation}
$A_{B B}$ and $A_{B^\ast B^\ast}$ are the prefactors of the incident $B B$ and $B^\ast B^\ast$ waves, respectively. $j_1(k_{B B} r)$ and $j_1(k_{B^\ast B^\ast} r)$ denote spherical Bessel functions representing the $L = 1$ contribution to these incident plane waves and $\chi_{B B}(r) / r$ and $\chi_{B^\ast B^\ast}(r) / r$ are the radial wave functions of the emergent $B B$ and $B^\ast B^\ast$ spherical waves. For large $r$, where $H_{\text{int}, S=0} = 0$, i.e.\ where the potential matrix (\ref{eq:potential-matrix-2x2}) vanishes, $\chi_\alpha(r) / r \propto h_1^{(1)}(k_\alpha r)$ with $\alpha = B B , B^\ast B^\ast$ and $h_1^{(1)}$ denoting a spherical Hankel function. This allows to define the $\mbox{T}$ matrix (see Eqs.\ (\ref{EQN008}) and (\ref{EQN009}) below).

Inserting the decomposition (\ref{EQN_phi_decomposition}) into Eq.\ (\ref{eq:spin0-equation}) and exploiting that incident plane waves are solutions of the free Schr\"odinger equation leads to
\begin{equation}
	\bigg(\begin{pmatrix} 2 m_B & 0 \\ 0 & 2 m_{B^\ast} \end{pmatrix} - \frac{1}{2 \mu} \bigg(\frac{d^2}{d r^2} - \frac{2}{r^2} \bigg) + H_{\text{int}, S=0} - E\bigg)
	\begin{pmatrix} \chi_{B B}(r) \\ \chi_{B^\ast B^\ast}(r) \end{pmatrix} =
	-H_{\text{int}, S=0}
	\begin{pmatrix} A_{B B} r j_1(k_{B B} r) \\ A_{B^\ast B^\ast} r j_1(k_{B^\ast B^\ast} r) \end{pmatrix} .
	\label{EQN013}
\end{equation}

%---------------------------------------------------------------------------------

\subsection{Boundary Conditions and Definition of the $\mbox{T}$ Matrix \label{sec:boundary-cond}}

As usual for radially symmetric problems in $3$ dimensional space, the wave functions $\chi_\alpha$ close to the origin must be proportional to $r^{L + 1}$, i.e.\ in our case
\begin{eqnarray}
	\label{EQN200} \chi_{\alpha}(r) \propto r^2  & & \text{for } r \rightarrow 0 .
\end{eqnarray}

For large $r$, where $H_{\text{int}, S=0} = 0$, the wave functions $\chi_\alpha$ describe exclusively emergent spherical waves and, thus, must be proportional to spherical Hankel functions. This leads to the definition of the $2 \times 2$ $\mbox{T}$ matrix,
\begin{equation}
	\mbox{T} = \begin{pmatrix}
	t_{B B ; B B} & t_{B B ; B^\ast B^\ast} \\
	t_{B^\ast B^\ast ; B B} & t_{B^\ast B^\ast ; B^\ast B^\ast}
	\end{pmatrix} ,
	\label{eq:t-matrix-definition}
\end{equation}
where the entries represent scattering amplitudes, which appear in the boundary conditions
\begin{eqnarray}
	\label{EQN008} \chi_\alpha(r) \propto i r t_{B B;\alpha} h_1^{(1)}(k_\alpha r) & & \text{for } r \rightarrow \infty \text{ and } (A_{B B} , A_{B^\ast B^\ast}) = (1 , 0) \\
	\label{EQN009} \chi_\alpha(r) \propto i r t_{B^* B^*;\alpha} h_1^{(1)}(k_\alpha r) & & \text{for } r \rightarrow \infty \text{ and } (A_{B B} , A_{B^\ast B^\ast}) = (0 , 1) .
\end{eqnarray}
Note that Eq.\ (\ref{EQN008}) describes a pure $B B$ incident wave, while Eq.\ (\ref{EQN009}) describes a pure $B^\ast B^\ast$ incident wave. To read off the $\mbox{T}$ matrix entries from Eqs.\ (\ref{EQN008}) and (\ref{EQN009}), one just has to solve Eq.\ (\ref{EQN013}) with boundary conditions (\ref{EQN200}) and (\ref{EQN008}) for $(A_{B B} , A_{B^\ast B^\ast}) = (1 , 0)$ as well as with boundary conditions (\ref{EQN200}) and (\ref{EQN009}) for $(A_{B B} , A_{B^\ast B^\ast}) = (0 , 1)$. How this can be done numerically is discussed in Section~\ref{sec:numerical-method}.

Poles of the $\mbox{T}$ matrix in the complex energy plane signal either bound states (for $\text{Re}(E_\text{pole}) < 2 m_B$ and $\text{Im}(E_\text{pole}) = 0$) or resonances (for $\text{Re}(E_\text{pole}) > 2 m_B$ and $\text{Im}(E_\text{pole}) < 0$). A pole implies that at least one of the eigenvalues of $\mbox{T}$ diverges or, equivalently, $1 / \text{det}(\mbox{T}) = 0$. We determine the poles of the $\mbox{T}$ matrix in Section~\ref{sec:pole-search} by numerically searching for roots of $1 / \text{det}(\mbox{T})$.

%---------------------------------------------------------------------------------

\subsection{\label{sec:numerical-method}Solving the Coupled-Channel Schr\"odinger Equation and Computing the $\mbox{T}$ Matrix for Given Energy $E$} 

To determine $t_{B B ; B B}$ and $t_{B B ; B^\ast B^\ast}$, the entries in the first row of the $\mbox{T}$ matrix \eqref{eq:t-matrix-definition} corresponding to an incident $B B$ wave, for given complex energy $E$, we proceed in the following way:
\begin{itemize}
\item[(1)] We solve the coupled-channel Schr\"odinger equation \eqref{EQN013} for coefficients $(A_{B B} , A_{B^* B^*}) = (1,0)$ using a standard fourth-order Runge-Kutta method. We start at very small $r$ and use three sets of initial conditions consistent with the boundary conditions \eqref{EQN200} to compute three functions:
\begin{itemize}
\item A solution of the inhomogeneous equation \eqref{EQN013}, denoted as $\vec{\chi}_\text{inhom}$, using $\vec{\chi} = (0 , 0)^{T} $ as initial conditions at small $r$ (here and in the following we use the notation $\vec{\chi} = (\chi_{B B} , \chi_{B^\ast B^\ast})^{T}$).
	
\item Two independent solutions of the homogeneous equation, i.e.\ Eq.\ \eqref{EQN013} with right hand side replaced by $0$, denoted as $\vec{\chi}_{0,a}$ and $\vec{\chi}_{0,b}$, using $\vec{\chi} = (r^2 , 0)^{T}$ and $\vec{\chi} = (0 , r^2)^{T}$ as initial conditions at small $r$, respectively.
\end{itemize}
The general numerical solution of the coupled-channel Schr\"odinger equation \eqref{EQN013} for given energy $E$ is then
\begin{eqnarray}
	\vec{\chi}_\text{numerical}(r) = \vec{\chi}_\text{inhom}(r) + a \vec{\chi}_{0,a}(r) + b \vec{\chi}_{0,b}(r)
\end{eqnarray}
with yet undetermined coefficients $a$ and $b$.

% *****

\item[(2)] In a second step we determine $a$ and $b$ according to the boundary conditions at $r \rightarrow \infty$, Eq.\ \eqref{EQN008}, i.e.\ we fix them in such a way, that $\vec{\chi}$ is exclusively an emergent speherical wave. To this end we divide the derivative of the boundary condition by the boundary condition and equate the result with the corresponding numerical expression evaluated at an arbitrary but sufficiently large $r = \tilde{r}$,
\begin{eqnarray}
	\frac{(d/dr) (r h_1^{(1)}(k_\alpha r))|_{r = \tilde{r}}}{\tilde{r} h_1^{(1)}(k_\alpha \tilde{r})} = \frac{(d/dr) \chi_{\text{numerical},\alpha}(r)|_{r = \tilde{r}}}{\chi_{\text{numerical},\alpha}(\tilde{r})} \quad , \quad \alpha = B B , B^\ast B^\ast .
\end{eqnarray}
These two equations are independent of the still unknown $\mbox{T}$ matrix elements and can be solved with respect to the two coefficients $a$ and $b$.

% *****

\item[(3)] Finally, we determine the $\mbox{T}$ matrix elements $t_{B B ; B B}$ and $t_{B B ; B^\ast B^\ast}$ from $\vec{\chi}_\text{numerical}(r)$ via
\begin{eqnarray}
t_{B B;\alpha} = \frac{\chi_{\text{numerical},\alpha}(\tilde{r})}{i \tilde{r} h_1^{(1)}(k_\alpha \tilde{r})} ,
\end{eqnarray}
where we have again used the boundary conditions at $r \rightarrow \infty$, Eq.\ \eqref{EQN008}.
\end{itemize}

The entries in the second row of the $\mbox{T}$ matrix \eqref{eq:t-matrix-definition}, $t_{B^\ast B^\ast ; B B}$ and $t_{B^\ast B^\ast ; B^\ast B^\ast}$, corresponding to an incident $B^\ast B^\ast$ wave, can be determined in the same way. One just has to use coefficients $(A_{B B} , A_{B^* B^*}) = (0,1)$ and replace the boundary conditions (\ref{EQN008}) by the boundary conditions (\ref{EQN009}).

%---------------------------------------------------------------------------------
%---------------------------------------------------------------------------------
%---------------------------------------------------------------------------------

\section{\label{SEC005}Numerical Results}

%---------------------------------------------------------------------------

\subsection{\label{SEC_parameters}Input Parameters and Error Analysis}

Unless explicitly stated otherwise, we use for the following numerical analysis a bottom quark mass $m_b = 4977 \, \text{MeV}$ taken from a quark model \cite{Godfrey:1985xj} and a meson mass splitting $m_{B^{*}} - m_{B} = 45 \, \text{MeV}$ as listed by the PDG \cite{ParticleDataGroup:2024cfk}. Moreover, as discussed in Section~\ref{sec2}, we use potential parameterizations (\ref{eq:0007}) with parameters $\alpha_5 = 0.34 \pm 0.03$,  $d_5 = 0.45^{+0.12}_{-0.10} \, \text{fm}$, $\alpha_j = -0.10 \pm 0.07$ and $d_j = (0.28 \pm 0.017) \text{fm}$ provided in Refs.\ \cite{Bicudo:2015kna,Bicudo:2016ooe}.

Uncertainties of numerical results in this work are estimated by propagation of uncertainties of the potential parameters.
For the parameters $\alpha_5 = 0.34 \pm 0.03$, $\alpha_j = -0.10 \pm 0.07$ and $d_j = (0.28 \pm 0.017) \text{fm}$, which have symmetric errors, we assume Gaussian distributions.
For the parameter $d_5 = 0.45^{+0.12}_{-0.10} \, \text{fm}$ we assume a skewed normal distribution with a positive skewness with parameters tuned in such a way, that the 16th and 84th percentiles correspond to the lower and upper errors of $d_5$.
From these distributions we draw 1000 random parameter samples $(\alpha_5,d_5,\alpha_j,d_j)$. Physical quantities like positions of $\mbox{T}$ matrix poles or branching ratios are computed for each of these samples. For some of these samples the imaginary part of the $\mbox{T}$ matrix pole is rather large and precisely locating the pole is numerically difficult. Moreover, a large imaginary part is equivalent to a large decay width, which implies that it is rather difficult or even impossible to experimentally detect the corresponding resonance.
We thus discard all poles with $\text{Im}(E_\text{pole}) < -200 \, \text{MeV}$ amounting to $\approx 14\%$ of the samples.
The uncertainties of pole positions and related quantities are then defined by the 16th and 84th percentiles on the remaining $\approx$ 86\% of the samples.

%---------------------------------------------------------------------------

\subsection{\label{sec:pole-search}Mass and Decay Width}

%---------------------------------------------------------------------------

\subsubsection{\label{SEC_solution_of_SE}Mass and Decay Width for Physical Parameters}

A pole of the $\mbox{T}$ matrix in the complex energy plane signals a tetraquark bound state (for $\text{Re}(E_\text{pole}) < 2 m_B$ and $\text{Im}(E_\text{pole}) = 0$) or a tetraquark resonance (for $\text{Re}(E_\text{pole}) > 2 m_B$ and $\text{Im}(E_\text{pole}) < 0$), where $E_\text{pole}$ denotes the pole position. The corresponding mass is
\begin{eqnarray}
m = \text{Re}(E_\text{pole})
\end{eqnarray}
and the decay width, in case of a resonance, is
\begin{eqnarray}
\Gamma = -2 \text{Im}(E_\text{pole}) .
\end{eqnarray}
We compute the $\mbox{T}$ matrix for given energy $E$ as discussed in Section~\ref{sec:numerical-method}. To locate possibly existing poles, we first scan the complex energy plane for singularities in $\text{det}(\mbox{T})$, i.e.\ for significantly enhanced $|\text{det}(\mbox{T})|$. This is done by sampling $\text{det}(\mbox{T})$ on a uniform grid in the region $1 \, \text{MeV} \leq \text{Re}(E - 2 m_B) \leq 120 \, \text{MeV}$ and $-100 \, \text{MeV} \leq \text{Im}(E - 2 m_B) \leq -1 \, \text{MeV}$ with spacing $1 \, \text{MeV}$ in both real and imaginary direction. When a singularity is indicated, we determine the corresponding pole position precisely by computing the nearby root of $1 / \text{det}(\mbox{T})$. For that we use Newton's method with an initial guess for $E$ provided by the energy scan.

\begin{figure}[htb]
	\centering
	\includegraphics[scale=0.36]{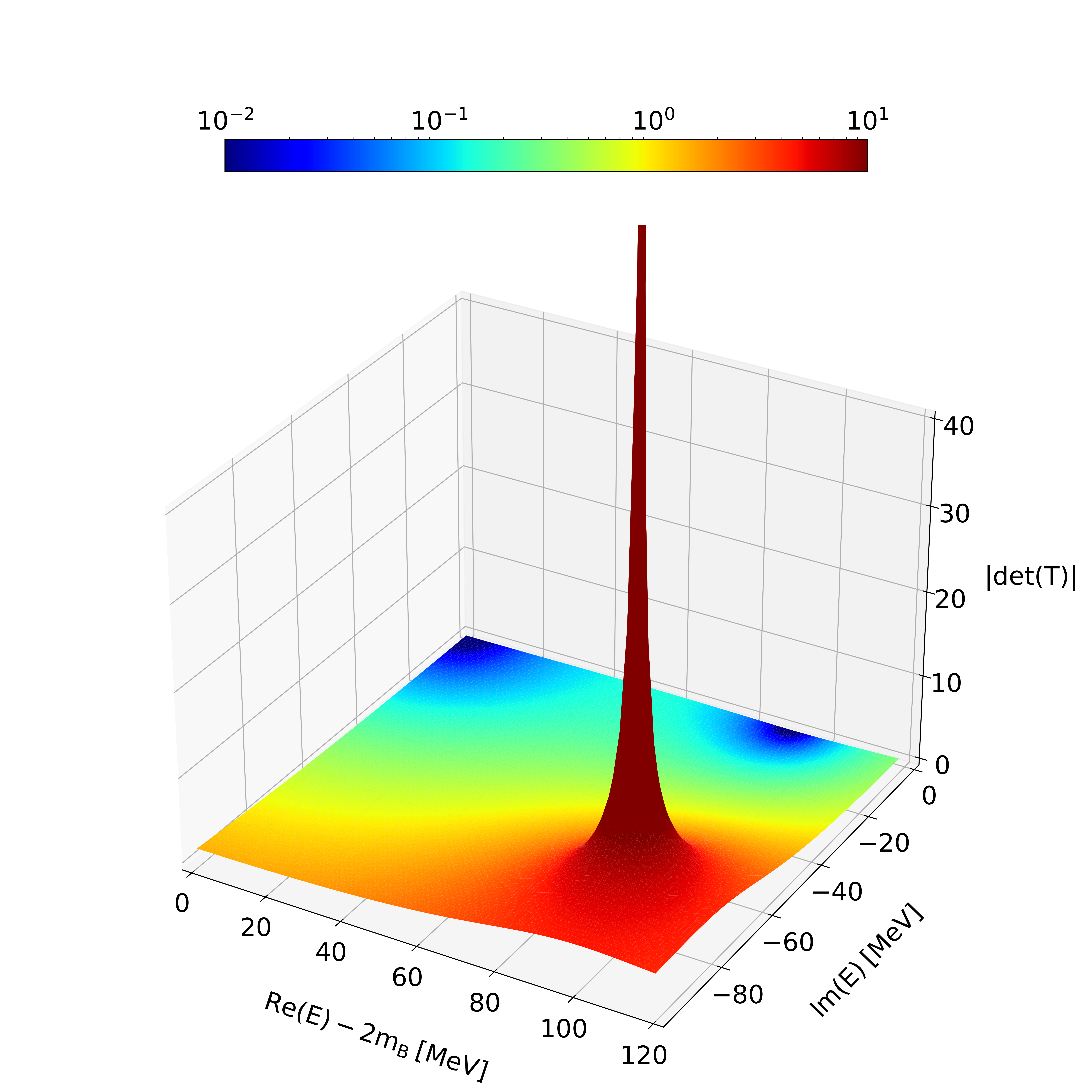}
	\caption{$|\text{det}(\mbox{T})|$ as function of the complex energy.}
	\label{fig:Tmatrix-poles-3D}
\end{figure}

For physical parameters, i.e.\ for physical bottom quark mass, meson mass splitting and potential parametrizations as listed in Section~\ref{SEC_parameters}, we find a pole with corresponding resonance mass and decay width
\begin{eqnarray}
\label{EQN_m_Gamma} m = 2 m_B + 94.0^{+1.3}_{-5.4} = 2 m_{B^\ast} + 4.0^{+1.3}_{-5.4} \, \text{MeV} \quad , \quad \Gamma = 140^{+86}_{-66} \, \text{MeV} .
\end{eqnarray}
This pole is illustrated in Figure~\ref{fig:Tmatrix-poles-3D}, where $|\text{det}(\mbox{T})|$ is plotted as a function of the complex energy $E$.

At first glance this result is surprising, because in our previous work \cite{Bicudo:2017szl}, based on a single-channel Schr\"odinger equation with heavy spin effects neglected, we predicted a resonance with mass $m_\text{single-channel} = 2 m_B + 17^{+4}_{-4} \, \text{MeV}$ slightly above the $B B$ threshold. Now, with our refined coupled-channel approach, the resonance mass is significantly larger, $m = 2 m_B + 94.0^{+1.3}_{-5.4} \, \text{MeV}$, i.e.\ slightly above the $B^\ast B^\ast$ threshold. This drastic shift in the resonance mass, when including heavy spin effects, is a direct consequence of the structure of the coupled-channel Schr\"odinger equation (\ref{EQN013}) and can be understood as follows.

The upper component of the 2-component wave function $\vec{\chi} = (\chi_{B B} , \chi_{B^\ast B^\ast})^{T}$ represents the $B B$ channel and the lower component the $B^\ast B^\ast$ channel, as indicated by the diagonal mass matrix $\text{diag}(2 m_B , 2 m_{B^\ast})$ in Eq.\ (\ref{EQN013}). Heavy spin effects are caused on the one hand by the mass splitting $m_{B^\ast} - m_B = 45  \, \text{MeV}$ and on the other hand by the $2\times 2$ potential matrix (\ref{eq:potential-matrix-2x2}).
The relevant potential for $B B$ is $(V_5(r) + 3 V_j(r)) / 4$ (the upper left entry of the potential matrix (\ref{eq:potential-matrix-2x2})).
The attractive contribution $V_5(r)$ is suppressed by the factor $1/4$ and there is an additional compensation of the attraction by the repulsive $V_j(r)$, which is enhanced by a relative factor of $3$. 
As a consequence, the net attractive force is negligible and a $B B$ resonance cannot be expected.
In contrast to that, the relevant potential for $B^\ast B^\ast$ is $(3 V_5(r) + V_j(r)) / 4$ (the lower right entry of the potential matrix (\ref{eq:potential-matrix-2x2})). This time, the attractive contribution $V_5(r)$ is only weakly suppressed by the factor $3/4$ and there is essentially no additional compensation of the attraction by the anyway rather weak repulsive $V_j(r)$. Consequently, a resonance with a sizable $B^\ast B^\ast$ component close to the $B^\ast B^\ast$ can be expected and is numerically found (see Eq.\ (\ref{EQN_m_Gamma}) and the discussion of branching ratios in Section~\ref{SEC_br}).

We note that an $I(J^P)=0(1^-)$ $\bar{b} \bar{b} u d$ tetraquark resonances was also predicted recently using a constituent quark model \cite{Meng:2021yjr}. The result from that reference, $m_{\text{\cite{Meng:2021yjr}}} = 10762(3) \, \text{MeV} \approx 2 m_B + 200 \, \text{MeV}$, is larger than our lattice QCD-based result, but still in crude qualitative agreement.

%---------------------------------------------------------------------------

\subsubsection{\label{SEC_theta}Varying the relative weights of $V_5(r)$ and $V_j(r)$ with respect to the $B B$ and the $B^\ast B^\ast$ Channel}

The $2 \times 2$ potential matrix \eqref{eq:potential-matrix-2x2} leads to a coupling of the $B B$ and the $B^\ast B^\ast$ channel.
To better understand the significant difference in the resonance mass, when comparing the coupled-channel result \eqref{EQN_m_Gamma} of this work to the single-channel result from Ref.\ \cite{Bicudo:2017szl}, we introduce an additional parameter $0 \leq \theta \leq \pi/2$ in the potential matrix \eqref{eq:potential-matrix-2x2},
\begin{eqnarray}
\nonumber & & \hspace{-0.7cm} H_{\text{int},S = 0} \ \ \rightarrow \ \ H_{\text{int},S = 0}(\theta) = R^T(\theta) \begin{pmatrix} V_5(r) & 0 \\ 0 & V_j(r) \end{pmatrix} R(\theta) = \\
\label{eq:general-potential-matrix} & & = \begin{pmatrix} \cos^2 \theta \, V_5(r) + \sin^2\theta \, V_j(r) & \sin \theta \cos \theta (V_5(r) - V_j(r)) \\ \sin \theta \cos \theta (V_5(r) - V_j(r)) & \sin^2 \theta \, V_5(r) + \cos^2 \theta \, V_j(r) \end{pmatrix} ,
\end{eqnarray}
where $R(\theta)$ is an ordinary rotation matrix,
\begin{equation}
R(\theta) = \begin{pmatrix} +\cos \theta & +\sin \theta \\ -\sin \theta & +\cos \theta \end{pmatrix} .
\end{equation}
For $\theta = 0$ the Schr\"odinger equation decouples into two independent equations, where the upper $B B$ equation is identical to the single-channel equation studied in Ref.\ \cite{Bicudo:2017szl}. $\theta = \pi/3$ corresponds to the coupled-channel equation \eqref{EQN013} derived in  Section~\ref{SEC_scattering} and solved in Section~\ref{SEC_solution_of_SE}, i.e.\ $H_{\text{int},S = 0}(\theta = \pi/3)$ is identical to the potential matrix \eqref{eq:potential-matrix-2x2}. For $\theta = \pi/2$ the Schr\"odinger equation decouples again into two independent equations, where this time the lower $B^\ast B^\ast$ equation is identical to the single-channel equation from Ref.\ \cite{Bicudo:2017szl} with exception of a constant energy shift by $2 (m_{B^\ast} - m_B) = 90 \, \text{MeV}$.
The parameter $\theta$, thus, allows to continuously rotate the attractive potential $V_5(r)$ from the $B B$ channel to the $B^\ast B^\ast$ channel and vice versa for the repulsive potential $V_j(r)$.

\begin{figure}[htb]
	\centering
	\includegraphics[scale=0.7]{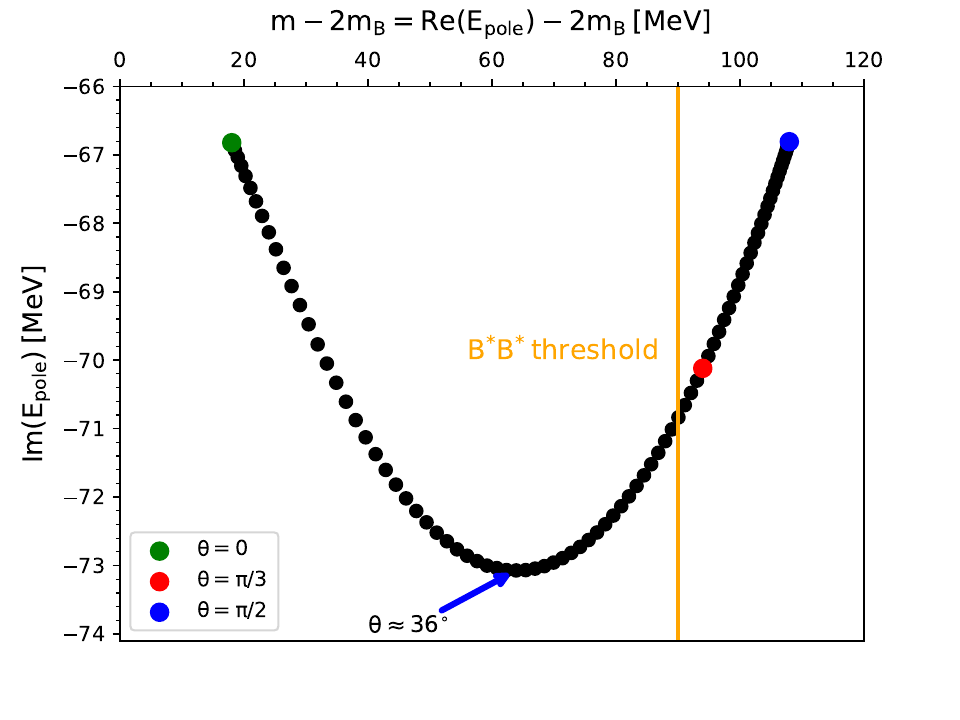}
	\caption{Trajectory of the $\mbox{T}$ matrix pole in the complex energy plane, when changing $\theta$ from $0$ to $\pi/2$.}
	\label{fig:Tmatrix-poles various theta}
\end{figure}

In Figure~\ref{fig:Tmatrix-poles various theta} we explore the effect of this rotation, i.e.\ we show the trajectory of the $\mbox{T}$ matrix pole in the complex energy plane, when changing $\theta$ from $0$ to $\pi/2$.
For $\theta = 0$ as well as for $\theta = \pi/2$ the result from Ref.\ \cite{Bicudo:2017szl} is reproduced \footnote{In the single channel case we find the resonance mass $18 \, \text{MeV}$ above threshold, i.e.\ there is a tiny difference to the $17 \, \text{MeV}$ quoted in Ref.\ \cite{Bicudo:2017szl}. The reason is that we use the $b$ quark mass in the kinetic term of the Schr\"odinger equation, while in Ref.\ \cite{Bicudo:2017szl} the $B$ meson mass was used.}, in the latter case shifted by $2 (m_{B^\ast} - m_B)$ as discussed in the previous paragraph.
It is, thus, not surprising that the resonance mass monotonically increases from $m \approx 2 m_B + 18 \, \text{MeV}$ at $\theta = 0$ to $m \approx 2 m_B + 108 \, \text{MeV} = 2 m_{B^\ast} + 18 \, \text{MeV}$ at $\theta = \pi/2$. At the physical point corresponding to $\theta = \pi/3$ the attractive potential is split between the $B B$ and the $B^\ast B^\ast$ channel in a 1:3 ratio, i.e.\ there is a sizable attraction between two $B^\ast$ mesons, but essentially no attraction between two $B$ mesons. This explains the location of the resonance mass \eqref{EQN_m_Gamma} close to the $B^\ast B^\ast$ threshold, as already discussed at the end of Section~\ref{SEC_solution_of_SE}.

%---------------------------------------------------------------------------

\subsubsection{\label{SEC_kappa}Varying the Heavy Quark Mass}

For sufficiently large heavy quark mass, the resonance found in Section~\ref{SEC_solution_of_SE} should turn into a bound state. We explore this by replacing
\begin{equation}
m_b \ \ \rightarrow \ \ \kappa m_b
\end{equation}
and continuously increasing $\kappa$, starting from $\kappa = 1$. We note that the meson mass difference $\Delta m = m_{B^\ast} -m_B$ strongly depends on the heavy quark mass, as can e.g.\ be seen by comparing experimentally known masses of the charm mesons $D$ and $D^\ast$ and of the bottom mesons $B$ and $B^\ast$. To account for this dependence, we use the leading order result from Heavy Quark Effective Theory \cite{Kilian_1993,Neubert_1994,neubert1996heavyquark},
\begin{equation}
\Delta m = \frac{m_{B^\ast}-m_B}{\kappa} = \frac{45 \, \text{MeV}}{\kappa} .
\end{equation}

\begin{figure}[htb]
	\centering
	\includegraphics[scale=0.6]{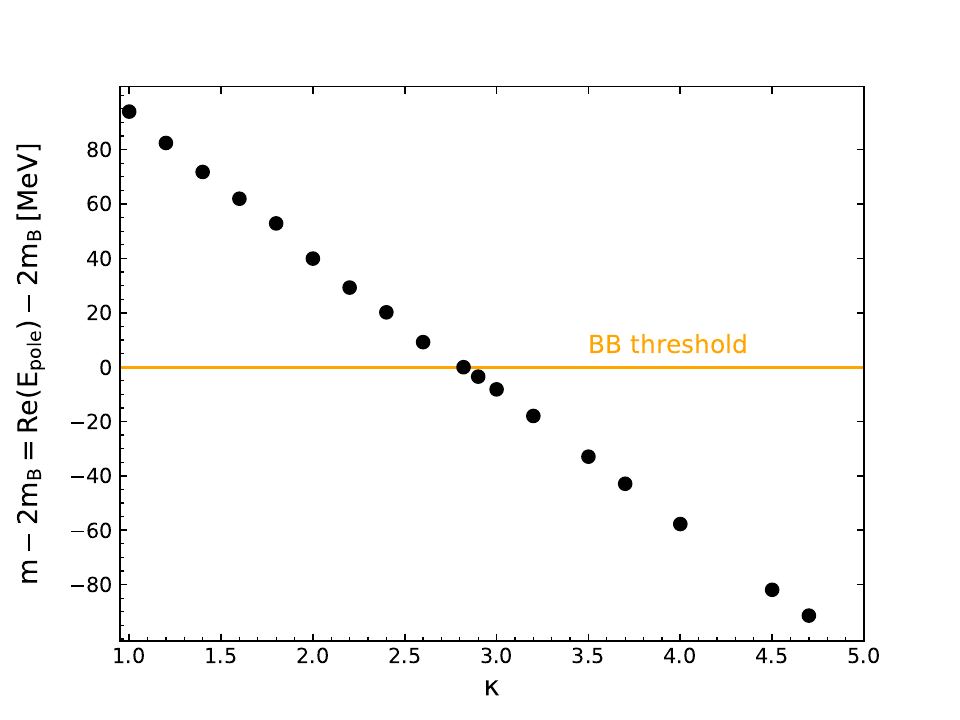}
	\caption{$m - 2 m_B = \text{Re}(E_\text{pole}) - 2 m_B$ as function of $\kappa$.}
	\label{fig:bound-state-various-kappa}
\end{figure}

In Figure~\ref{fig:bound-state-various-kappa} we show $m - 2 m_B = \text{Re}(E_\text{pole}) - 2 m_B$ as function of $\kappa$. When increasing $\kappa$, starting from $\kappa = 1$, the resonance moves closer to the $B B$ threshold, until at $\kappa \approx 2.82$ it becomes a bound state. The dependence of the resonance or bound state mass on $\kappa$ exhibits an almost linear behavior.

\begin{figure}[H]
	\centering
	\includegraphics[scale=0.65]{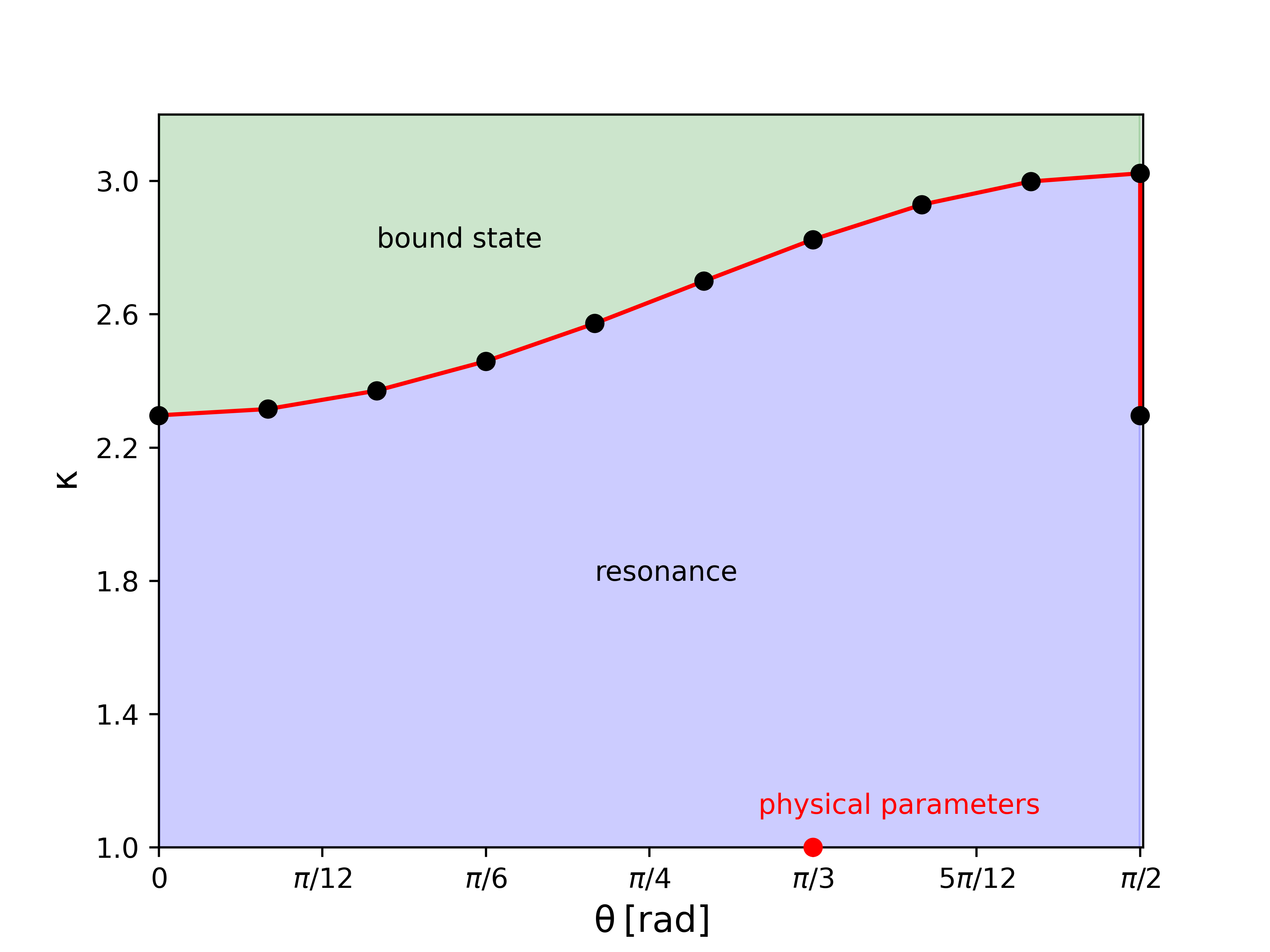}
	\caption{Existence of a resonance or bound state in the $\theta$-$\kappa$ plane (see Section~\ref{SEC_theta} and Section~\ref{SEC_kappa} for detailed explanations of these parameters). Note that at $\theta = \pi/2$ there is a discontinuity in the curve separating the two regions, because the $2 \times 2$ Schr\"odinger equation decouples into two independent equations and the lowest threshold suddenly changes from $2 m_B$ to $2 m_{B^\ast}$.}
	\label{fig:t-matrix-various-kappa-theta}
\end{figure}

In Figure~\ref{fig:t-matrix-various-kappa-theta} we visualize, how the existence of a resonance or bound state depends on $\theta$ (i.e.\ the splitting of the potentials $V_5(r)$ and $V_j(r)$ between the $B B$ and the $B^\ast B^\ast$ channel) and on $\kappa$ (i.e.\ the heavy quark mass). The black dots (connected by straight red lines to guide the eye) separate a resonance (blue region) from a bound state (green region).

%---------------------------------------------------------------------------

\subsection{\label{SEC_br}Branching Ratios}

In addition to the mass and the decay width of the $I(J^P) = 0(1^-)$ $\bar b \bar b u d$ tetraquark resonance, it is also of interest to study its decay channels. The resonance can decay into a $B B$ meson pair or into a $B^\ast B^\ast$ meson pair.
In experiments decay probabilities to different channels $\alpha$ are referred to as branching ratios $\text{BR}_{\alpha}$.
Theoretically, branching ratios can be approximated by the residues of the diagonal entries of the $\mbox{T}$ matrix at the pole energy (see e.g.\ Ref.\ \cite{Burkert:2022bqo}), in our case
\begin{equation}
\text{BR}_\alpha = \frac{|\text{Res}(t_{\alpha;\alpha})|}{|\mathrm{Res}(t_{B B;B B})| + |\mathrm{Res}(t_{B^\ast B^\ast;B^\ast B^\ast})|} \quad , \quad \alpha = B B , B^\ast B^\ast .
\label{eq:Branching-ratios}
\end{equation}

To compute the residues we use the residue theorem
\begin{equation}
\mathrm{Res}(t_{\alpha;\alpha}) = \frac{1}{2\pi i} \oint \limits_{\mathcal{C}} dE \, t_{\alpha;\alpha}(E) ,
\label{EQN_ResidueTheorem}
\end{equation}
where $\mathcal{C}$ is a closed contour in the complex energy plane encircling the $\mbox{T}$ matrix pole at position $E_\text{pole}$. We choose a circle with radius $r = 1 \, \text{MeV}$ around the pole as contour $\mathcal{C}$ parameterized by
\begin{equation}
E(\lambda) = E_\text{pole} + r e^{2 \pi i \lambda}
\label{eq:contour-param}
\end{equation}
with $0 \leq \lambda \leq 1$. The integral (\ref{EQN_ResidueTheorem}) can be solved numerically in a straightforward way. We verified the stability of the resulting residues and branching ratios by varying the number of integration points and the radius $r$.

\begin{figure}[H]
	\centering
	\includegraphics[scale=0.6]{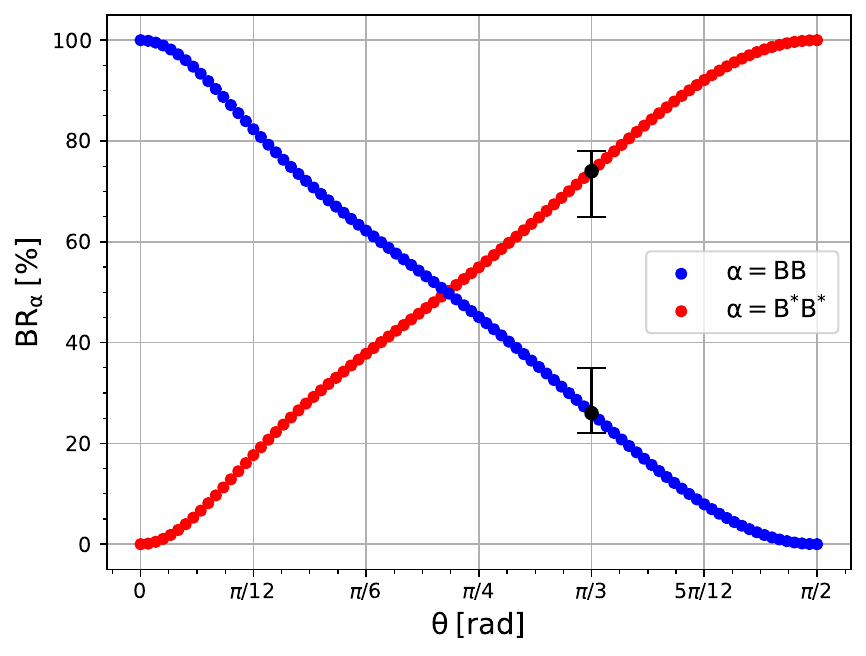}
	\caption{Branching ratios $\text{BR}_{B B}$ and $\text{BR}_{B^\ast B^\ast} = 1 - \text{BR}_{B B}$ as functions of the angle $\theta$ defined in Section~\ref{SEC_theta}.}
	\label{fig:branching-ratios}
\end{figure}

In Figure~\ref{fig:branching-ratios} we show the branching ratios $\text{BR}_{B B}$ and $\text{BR}_{B^\ast B^\ast} = 1 - \text{BR}_{B B}$ as functions of the angle $\theta$ defined in Section~\ref{SEC_theta}.
For the physical value $\theta = \pi/3$ we find
\begin{equation}
\label{EQN_BR} \text{BR}_{B B} = 26^{+9}_{-4} \% \quad , \quad \text{BR}_{B^\ast B^\ast} = 74^{+4}_{-9} \%
\end{equation}
indicating a significantly favored decay of the tetraquark resonance to $B^\ast B^\ast$. This is consistent with our results from Section~\ref{SEC_solution_of_SE}, where we found a resonance mass close to the $B^\ast B^\ast$ threshold suggesting that the resonance is mainly composed of a $B^\ast B^\ast$ pair and that $\text{BR}_{B^\ast B^\ast} \gg \text{BR}_{B B}$.

Decreasing the angle $\theta$ shifts the attractive potential $V_5(r)$ towards the $B B$ channel, which is reflected in an increasing branching ratio $\text{BR}_{B B}$. At $\theta = 0$ the Schr\"odinger equation (\ref{EQN013}) decouples into a $B B$ and a $B^\ast B^\ast$ equation, as discussed in Section~\ref{SEC_theta}. The resonance is then obtained from the $B B$ equation (which was already studied in Ref.\ \cite{Bicudo:2017szl}) and, consequently, $\text{BR}_{B B} = 100 \%$. Similarly, increasing the angle $\theta$ shifts the attractive potential $V_5(r)$ towards the $B^\ast B^\ast$ channel, leading to an increasing branching ratio $\text{BR}_{B^\ast B^\ast}$, where $\text{BR}_{B^\ast B^\ast} = 100 \%$ at $\theta = \pi/2$.

%---------------------------------------------------------------------------
%---------------------------------------------------------------------------
%---------------------------------------------------------------------------

\section{\label{SEC_conclusions}Conclusions and Outlook}

In this work we predicted a $\bar{b} \bar{b} u d$ tetraquark resonance with quantum numbers $I(J^P) = 0(1^-)$ slightly above the $B^\ast B^\ast$ threshold using antistatic-antistatic potentials computed with lattice QCD and the Born-Oppenheimer approximation.
The mass and decay width of this resonance are $m = 2 m_B + 94.0^{+1.3}_{-5.4} \, \text{MeV} = 2 m_{B^\ast} + 4.0^{+1.3}_{-5.4} \, \text{MeV}$ and $\Gamma = 140^{+86}_{-66} \, \text{MeV}$.
On a technical level, we significantly refined our single-channel approach from Ref.\ \cite{Bicudo:2017szl}, where we did not distinguish between $B$ and $B^\ast$ mesons, which are degenerate in the static limit. Now we incorporated their mass difference of around $45 \, \text{MeV}$ by setting up a coupled-channel Schr\"odinger equation containing not only an attractive potential $V_5(r)$ (already used in Ref.\ \cite{Bicudo:2017szl}), but also a repulsive potential $V_j(r)$.

In addition to the numerical prediction of the mass and decay width, our formalism also led to a solid physical understanding, why there is a $\bar{b} \bar{b} u d$ tetraquark resonance close to the $B^\ast B^\ast$ threshold, but not in the region of the $B B$ threshold, as naively expected from our previous work \cite{Bicudo:2017szl}. The reason is that the attractive potential $V_5(r)$ dominates the $B^\ast B^\ast$ channel, but is strongly suppressed in the $B B$ channel, whereas the situation is reversed for the repulsive potential $V_j(r)$.
This theoretical result is consistent with and supported by further numerical results on
branching ratios, which indicate that a decay of the $I(J^P) = 0(1^-)$ $\bar{b} \bar{b} u d$ tetraquark resonance is around three times more likely to a $B^\ast B^\ast$ pair than to a $B B$ pair.

The Born-Oppenheimer approach and the antistatic-antistatic lattice QCD potentials, on which our work is based, introduce systematic errors, which are difficult to quantify precisely. As a crude estimate we compare results obtained for a related QCD-stable $\bar{b} \bar{b} u d$ tetraquark with quantum numbers $I(J^P) = 0(1^+)$. The Born-Oppenheimer approach using the same lattice QCD potentials and the same formalism to derive a coupled-channel Schr\"odinger equation leads to a binding energy of $59_{-38}^{+30} \, \text{MeV}$ \cite{Bicudo:2016ooe}. This binding energy has, meanwhile, been computed within full lattice QCD by several independent groups, where their results point towards a binding energy around $100 \, \text{MeV}$ \cite{Francis:2016hui,Junnarkar:2018twb,Leskovec:2019ioa,Mohanta:2020eed,Hudspith:2023loy,Aoki:2023nzp,Alexandrou:2024iwi}. In view of this comparison it seems likely that an $I(J^P) = 0(1^-)$ $\bar{b} \bar{b} u d$ tetraquark resonance exists, since the Born-Oppenheimer approach with the used potentials leads to results in fair agreement with full lattice QCD predictions, even having a slight tendency of underestimating the binding.

This work is also intendend as an exploratory and preparatory investigation for a future full lattice QCD investigation of the predicted $I(J^P) = 0(1^-)$ $\bar{b} \bar{b} u d$ tetraquark resonance. The most important findings and main takeaways for such a full lattice QCD investigation are the following:
\begin{itemize}
\item[(i)] The existence of a tetraquark resonance can be expected close to the $B^\ast B^\ast$ threshold, i.e.\ significantly above the $B B$ threshold. 

\item[(ii)] Because of (i) and since the $\bar{b} \bar{b} u d$ tetraquark resonance predominantly decays to a $B^\ast B^\ast$ pair, it will be imperative to use scattering interpolating operators of both $B B$ and $B^\ast B^\ast$ type, to reliably extract finite volume energy levels up to the region of the $B^\ast B^\ast$ threshold.

\item[(iii)] Because of (i) and (ii), a standard single-channel scattering analysis of finite volume energy levels is not sufficient. One rather has to resort to technically more difficult coupled-channel finite volume scattering techniques (see e.g.\ Ref.\ \cite{Hansen:2012tf}).
\end{itemize}
In recent work we have prepared and tested a lattice QCD setup with $\bar{b} \bar{b} u d$ scattering operators both at the source and at the sink of correlation functions \cite{Alexandrou:2024iwi}. This setup is perfectly suited for an implementation of the $B B$ and $B^\ast B^\ast$ scattering operators mentioned in (ii). We plan to continue our work in this direction in the near future with the main goal to arrive at a rigorous full lattice QCD prediction of the $I(J^P) = 0(1^-)$ $\bar{b} \bar{b} u d$ tetraquark resonance using finite-volume methods as referenced in (iii).

% ********************
% ********************
% ********************

\section*{Acknowledgements}

We thank André Zimermmane-Santos for collaboration on earlier related work.
We acknowledge useful discussions with Pedro Bicudo, Lasse Müller and Martin Pflaumer.

J.H.\ acknowledges support by a ``Rolf and Edith Sandvoss Stipendium''.
M.W.\ acknowledges support by the Deutsche Forschungsgemeinschaft (DFG, German Research Foundation) -- project number 457742095.
M.W.\ acknowledges support by the Heisenberg Programme of the Deutsche Forschungsgemeinschaft (DFG, German Research Foundation) -- project number 399217702.

Calculations on the GOETHE-NHR and on the FUCHS-CSC  high-performance computers of the Frankfurt University were conducted for this research. We would like to thank HPC-Hessen, funded by the State Ministry of Higher Education, Research and the Arts, for programming advice.

% ********************
% ********************
% ********************
%\printbibliography
\bibliographystyle{utphys}
\bibliography{References}

% ********************
% ********************
% ********************

\end{document}